\documentstyle[12pt]{article}

\newcommand{\ie}{{\it i.e.}\ }

\newcommand{\lsim}{\buildrel < \over {_\sim}}

\newcommand{\ccbar}{c\bar{c}}
\newcommand{\QQbar}{Q\bar{Q}}
\newcommand{\qqbar}{q\bar{q}}
\newcommand{\ppbar}{p\bar{p}}

\newcommand{\into}{\rightarrow}
\newcommand{\jp}{$J/\psi$}
\newcommand{\jps}{$J/\psi$\ }
\newcommand{\as}{$\alpha_s$}

\newcommand{\als}{\alpha_s}

\newcommand{\half}{\frac{1}{2}}
\newcommand{\beq}{\begin{equation}}
\newcommand{\eeq}{\end{equation}}
\newcommand{\beqa}{\begin{eqnarray}}
\newcommand{\eeqa}{\end{eqnarray}}

\newcommand{\PLB}[3]{\mbox{}Phys. Lett. {\bf B{#1}}, {#2} ({#3})}
\newcommand{\NPB}[3]{\mbox{}Nucl. Phys. {\bf B{#1}}, {#2} ({#3})}
\newcommand{\PR}[3]{\mbox{}Phys. Rev. {\bf {#1}}, {#2} ({#3})}
\newcommand{\PRL}[3]{\mbox{}Phys. Rev. Lett. {\bf {#1}}, {#2} ({#3})}
\newcommand{\PRD}[3]{\mbox{}Phys. Rev. {\bf D{#1}}, {#2} ({#3})}
\newcommand{\ZPC}[3]{\mbox{}Z. Phys. {\bf C{#1}}, {#2} ({#3})}

\newcommand{\etal}{{\em et al.}}

\newcommand{\octet}[1]{c\bar{c} \left[ \,{#1}^{(8)} \right]}

\newcommand{\octetqn}[2]{c\bar{c} \left[ \,{#1}^{(8)} ; {#2} \right]}

\newcommand{\process}[2]{{#1} \rightarrow {#2} \rightarrow \psi' + X}

\newcommand{\square}[2]{\left| A \left( {#1} \rightarrow {#2}
\rightarrow \psi'(\lambda) + X \right) \right| ^2}
\newcommand{\element}[1]{<0|{\cal O}^{\psi'}_8 (\,{#1} )|0>}

\newcommand{\trace}[2]{\, {\rm Tr} \left. \left[
{\cal O}_{{#1}\rightarrow c\bar{c}}^{AB}(P,q) P_{{#2}}(P,q)
\right] \right| _{q=0}}

\begin{document}

\setlength{\baselineskip}{7mm}

\thispagestyle{empty}
\begin{flushright}
   \vbox{\baselineskip 12.5pt plus 1pt minus 1pt
         SLAC-PUB-95-6931 \\
         HU-TFT-95-35 \\
         June 1995 }
\end{flushright}

\renewcommand{\thefootnote}{\fnsymbol{footnote}}
\bigskip
\begin{center}
{\Large \bf Color-Octet $\psi'$ Production at Low $p_\perp$}

\vskip 1\baselineskip

Wai-Keung Tang \footnote{%
Work supported in part by Department of Energy contract DE--AC03-
-76SF00515 and DE--AC02--76ER03069. }   \\
{\normalsize \em Stanford Linear Accelerator Center } \\
{\normalsize \em Stanford University, Stanford, CA 94309}

\vskip 1\baselineskip

M. V\"anttinen\footnote
{Work supported by the Academy of Finland
under project number 8579.}\footnote{Address after September 1, 1995:
NORDITA, Copenhagen.} \\
{\normalsize \em Research Institute for Theoretical Physics } \\
{\normalsize \em P. O. Box 9, FIN-00014 University of Helsinki, Finland} \\

\vskip 1\baselineskip

\end{center}

\medskip

\renewcommand{\thefootnote}{\arabic{footnote}}
\setcounter{footnote}{0}

\begin{center}
{\bf ABSTRACT}
\end{center}
\bigskip

\vbox{\baselineskip 14pt
\noindent
We study contributions from color-octet quarkonium formation mechanisms
to $p_\perp$-integrated $\psi'$ production cross sections in pion-nucleon
reactions. The observed polarization of the $\psi'$ is not reproduced by
the lowest-order leading-twist color-singlet and color-octet mechanisms.
This suggests that there are
important quarkonium production mechanisms beyond leading twist.
}

\bigskip
\begin{center}
{\it Submitted to Physical Review Letters}
\end{center}
\newpage

The production of heavy quarkonium has been widely studied using
perturbative QCD. Due to the large masses of the $c$ and $b$
quarks, quarkonium production amplitudes can be factorized as
products of perturbative $\QQbar$ production amplitudes and
non-perturbative quarkonium formation amplitudes. In the
conventional model of hadroproduction \cite{BaierZPC19},
a $\QQbar$ pair with the appropriate quantum numbers is
produced in the collision of two partons, \ie at leading
twist, and the formation process is represented by a color-singlet
Schr\"odinger wavefunction.

The CDF collaboration has recently reported several
measurements of quarkonium production in $\ppbar$ collisions at
$\sqrt{s} = $ 1.8 TeV \cite{CDFcharmonium,CDFbottomonium} which
contradict the predictions of the conventional model. Possible
theoretical explanations of these experimental results include parton
fragmentation into quarkonium, which was first studied in
perturbative QCD by Braaten and Yuan \cite{BraatenYuan}.
Unlike the conventional parton-parton fusion mechanisms \cite{BaierZPC19},
fragmentation is consistent with the approximate
\mbox{$1/p_\perp^4$} shape of the cross sections
\mbox{$d\sigma/dp_\perp(\ppbar \into \psi(1S,2S)+X)$} at large $p_\perp$.
However, if the quarkonia are treated as color-singlet $\QQbar$
systems, the normalization of the cross sections still falls short
of the data by as much as a factor of 5 for $J/\psi(1S)$ production
and a factor of 30 for $\psi'(2S)$ production.

A number of possible resolutions have been suggested to explain the
observed $\psi'$ surplus \cite{CTW,Close,CW,BraatenNUHEPTH9426}. In
particular, Braaten and Fleming have proposed
that non-perturbative transitions between color-octet and color-singlet
$\ccbar$ states be included in the fragmentation calculation
\cite{BraatenNUHEPTH9426}.
The inclusion of color-octet matrix elements is attractive both from the
theoretical and phenomenological point of view. Theoretically, octet
contributions are an essential part of the perturbative picture
\cite{peskin,Bodwin}. Phenomenologically, octet contributions can boost
the cross section of heavy quark production, not just at
high $p_{\perp}$, but also at low $p_{\perp}$, where large discrepancies
are also observed \cite{VHBT}.

In the mechanism proposed by Braaten and Fleming, a gluon first produces a
color-octet $\ccbar$ pair in a $^3 S_1$ angular momentum state.
The color-octet pair then evolves into a $\psi'$
by a non-perturbative double chromoelectric dipole transition
described by the matrix element $<0|{\cal O}_8^{\psi'}(^3 S_1)|0>$.
This process is illustrated in Fig. 1, where the non-perturbative
transition is represented by the blob.
The non-perturbative matrix element is related to the weight of the
$|\ccbar gg>$ Fock state in the $\psi'$ wavefunction. Because the
matrix element is not constrained by other existing data,
it can be adjusted to fit the CDF data.
The value obtained in Ref. \cite{BraatenNUHEPTH9426}
is
$$
  <0|{\cal O}_8^{\psi'}(^3 S_1)|0> = 0.0042 \; {\rm GeV}^3.
$$
Cho and Leibovich \cite{ChoLeibovich}, on the other hand,
have combined color-octet transitions with a
formula that
smoothly interpolates between parton-parton fusion mechanisms and
fragmentation mechanisms of $\QQbar$ production. They find that
$$
  <0|{\cal O}_8^{\psi'}(^3 S_1)|0> = 0.0073 \; {\rm GeV}^3.
$$

Color-octet mechanisms were originally introduced
as part of a rigorous treatment of P-wave charmonium decays
\cite{Bodwin}. In the color-singlet model some of these decay
widths, as well as the corresponding leading-twist P-wave
production cross sections, are infrared divergent.
The divergences can be absorbed into the non-perturbative color-octet
matrix elements. In S-wave production and decay, such
divergences do not appear. Color-octet production mechanisms
nevertheless provide an attractive explanation of the
wrong normalization of quarkonium production. Note that CDF
have also observed a surplus of S-wave bottomonium
\cite{CDFbottomonium} with respect to the
color-singlet parton-parton fusion
predictions. The bottomonium surplus appears also at
\mbox{$p_\perp \lsim m_b$},
where parton-parton fusion
mechanisms are expected to dominate over fragmentation mechanisms.

The predictions of the color-singlet parton-parton fusion model
disagree with fixed-target hadroproduction data as well. In a recent paper
on charmonium production in pion-nucleon reactions \cite{VHBT},
we pointed out that the model fails to reproduce the
relative production rates of the \jp, $\chi_1$ and $\chi_2$ states
\cite{AntoniazziPRL70} and the polarization of the \jps
\cite{Biino} and $\psi'$ \cite{Akerlof}.
We interpreted these discrepancies as evidence for important higher-twist
mechanisms of charmonium production. It is, however, necessary
to check whether the data could be reproduced within the leading-twist
picture by including the color-octet transitions. At the same time,
such an analysis could provide an independent determination of
the color-octet matrix elements whose values have previously been
extracted from CDF data.

In this letter, we study the $p_\perp$-integrated $\psi'$ production
cross section in pion-nucleon collisions. The analysis also applies
to the direct component of \jps production. Total \jps production
is a somewhat more complicated process because there are significant
contributions from the decays of the $\chi_J$ and $\psi'$ states
\cite{AntoniazziPRL70}.

The polarization of the $\psi'$ gives important information on its
production mechanisms.
Measurements of polarization actually indicate
that there must be other
production mechanisms in addition to the leading-twist
color-singlet and color-octet mechanisms.

At leading order in \as\ and up to next-to-leading order in $v^2/c^2$,
the subprocesses for $\psi'$ production through color-octet
intermediate states are
\beqa
  \process{gg}{\octet{^3P_J}}, \label{Pwaveoctet} \\
  \process{\qqbar}{\octet{^3S_1}}, \label{vectoroctet} \\
  \process{gg}{\octet{^1S_0}}. \label{etaoctet}
\eeqa
These are illustrated in Fig. 2. The process
\mbox{$\process{\qqbar}{\octet{^3P_J}}$} is of higher order in
$v^2/c^2$ since the lowest-order non-perturbative transition
(single chromoelectric dipole) is forbidden by charge conjugation.
We also find that the subprocess amplitude
\mbox{$ A \left( gg \into \octet{^3S_1} \right)$}
vanishes in the leading order in \as. Note that charge conjugation
or Yang's theorem \cite{Yang} do not require this amplitude to vanish
because the $\ccbar$ pair is not in a color-singlet state.

The notation in eq. (1) includes a total angular momentum quantum number $J$.
However, we assume that the color-octet states
with different $J$ do not propagate as resonances of different masses.
Therefore, unlike in the case of color-singlet P-wave states \cite{VHBT}, we
can neglect the $JJ_z$ coupling and work in the $L_z S_z$ basis
throughout.

The polarization of the final-state $\psi'$ is measured by
the polar-angle distribution, $1 + \alpha\cos^2\theta$, of its decay
dileptons in their rest frame. The parameter $\alpha$ in the
Gottfried-Jackson frame angular distribution is related to the
polarized $\psi'$ production cross sections in the following way:
\beq
  \alpha = \frac{d\sigma(\lambda=1) - d\sigma(\lambda=0)}{
           d\sigma(\lambda=1) + d\sigma(\lambda=0)},
\eeq
where $\lambda$ is the helicity of the $\psi'$. Since
the transverse momenta of the initial partons and the momenta of the
soft gluons emitted in the non-perturbative transition are small and
can be neglected, the
helicity of the $\psi'$ is equal to the $z$ component of its spin.
It is determined by the perturbative dynamics of the subprocesses
$\qqbar,gg \into \ccbar$ and by the heavy quark spin symmetry of
the non-perturbative transition. In the process (\ref{Pwaveoctet}),
the non-perturbative transition is a chromoelectric dipole transition.
In (\ref{vectoroctet}), it is a double chromoelectric dipole transition
and in (\ref{etaoctet}), a chromomagnetic dipole transition.
In the chromoelectric dipole transitions, the helicities of the heavy quark
and antiquark are not flipped, whereas in the  chromomagnetic dipole
transition one of the helicities is flipped.
In processes
(\ref{Pwaveoctet}) and (\ref{vectoroctet}),
the helicity $\lambda$ of the $\psi'$
therefore equals the $z$ component of the spin of the color-octet
$\ccbar$ state. In process (\ref{etaoctet}), the total spin
of the quark-antiquark system changes from $S = 0, \; S_z = 0$ in the
color-octet state to  $S(\psi') = 1, \; S_z(\psi') = \lambda = \pm 1$
in the final state.

In terms of the matrix elements of non-relativistic QCD, the squares
of the amplitudes of processes
(\ref{Pwaveoctet}), (\ref{vectoroctet}) and (\ref{etaoctet}) are
\beqa
  \lefteqn{\square{gg}{\octet{^3P_J}}} \nonumber \\
  & = & \frac{1}{24m_c} \element{^3P_J}
        \delta_{aa'} \delta_{\lambda S_z} \delta_{\lambda S_z'}
        \delta_{L_z L_z'} \nonumber \\
  &   & \times A \left( gg \into \octetqn{^3P_J}{a,S_z,L_z} \right)
        A^* \left( gg \into \octetqn{^3P_J}{a',S_z',L_z'} \right) , \\
  \lefteqn{\square{\qqbar}{\octet{^3S_1}}} \nonumber \\
  & = & \frac{1}{24m_c} \element{^3S_1}
        \delta_{aa'} \delta_{\lambda S_z} \delta_{\lambda S_z'}
        \nonumber \\
  &   & \times A \left( \qqbar \into \octetqn{^3S_1}{a,S_z} \right)
        A^* \left( \qqbar \into \octetqn{^3S_1}{a',S_z'} \right) , \\
  \lefteqn{\square{gg}{\octet{^1S_0}}} \nonumber \\
  & = & \frac{1}{24m_c} \element{^1S_0}
        \delta_{aa'} \half (1-\delta_{\lambda 0})
        \nonumber \\
  &   & \times A \left( gg \into \octetqn{^1S_0}{a} \right)
        A^* \left( gg \into \octetqn{^1S_0}{a'} \right) ,
\eeqa
where $a,a',S_z,S_z',L_z,L_z'$ are the color quantum numbers and
angular momentum components of the color-octet intermediate states.
The perturbative amplitudes are
\beqa
  A \left( gg \into \octetqn{^3P_J}{a,S_z,L_z} \right)
    & = & \sqrt{2} \; T^a_{AB} \epsilon^{*\alpha}(L_z)
          \nonumber \\
    &   & \times
          \frac{\partial}{\partial q^\alpha} \trace{gg}{1S_z}, \\
  A \left( \qqbar \into \octetqn{^3S_1}{a,S_z} \right)
    & = & \sqrt{2} \; T^a_{AB} \trace{\qqbar}{1S_z}, \\
  A \left( gg \into \octetqn{^1S_0}{a} \right)
    & = & \sqrt{2} \; T^a_{AB} \trace{gg}{00},
\eeqa
where $P$ is the total four-momentum and $2q$ the relative
four-momentum of the $\ccbar$ pair, ${\cal O}_{ij \into \ccbar}^{AB}$
is the perturbative amplitude for the production of a $\ccbar$ pair
with color indices $A,B$ (the heavy quark legs are truncated),
$\epsilon$ is a spin-1 polarization four-vector, and $P_{1S_z}, P_{00}$
are the covariant spin projection operators of Ref. \cite{ChoLeibovich}.
Taking the four-momentum of the $\psi'$ equal to $P$, the $\psi'$
production cross section is
\beqa
  \sigma(\pi N \into \psi'(\lambda) + X)
  & = & \sum_{ij} \int dx_1 dx_2 \; f_{i/\pi}(x_1) \; f_{j/N}(x_2)
        \nonumber \\
  &   & \times \delta \left( 1 - \frac{M^2}{x_1 x_2 s} \right)
        \frac{\pi}{M^4}
        \overline{ | A(ij \into \psi'(\lambda) + X) |^2} \nonumber \\
  & \equiv & \sum_{ij} \Phi_{ij/\pi N} (M^2/s)
        \frac{\pi}{M^4}
        \overline{ | A(ij \into \psi'(\lambda) + X) |^2},
\eeqa
where we neglected factorization-scale dependence.
The contribution from the color-octet subprocesses
(\ref{Pwaveoctet}), (\ref{vectoroctet}) and
(\ref{etaoctet}) simplifies to
\beqa
  \lefteqn{\sigma_{\rm octet} (\pi N \into \psi'(\lambda)
   + X)} \nonumber \\
  & = & \frac{\pi}{M^4} \left\{
        \Phi_{gg/\pi N}(M^2/s) \left[
        \frac{10 \pi^2 \als^2 m_c}{9 M^4}
        \element{^3P_J}
        (3 - 2\delta_{\lambda 0})
        \right. \right. \nonumber \\
  &   & \hspace{40mm} \mbox{} + \left.
        \frac{5 \pi^2 \als^2}{144 m_c} \element{^1S_0}
        (1 - \delta_{\lambda 0}) \right] \nonumber \\
  &   & \mbox{} + \sum_q \left[
        \Phi_{q\bar{q}/\pi N}(M^2/s) + \Phi_{\bar{q}q/\pi N}(M^2/s)
        \right] \nonumber \\
  &   & \hspace{20mm} \times \left.
        \frac{32 \pi^2 \als^2 m_c}{54 M^2}
        \element{^3S_1}
        (1 - \delta_{\lambda 0})
        \right\} .
  \label{octetcontribution}
\eeqa
The difference between the measured cross section
\cite{AntoniazziPRL70}
$$\sigma (\pi N \into \psi' + X; x_F>0) = 25 \; {\rm nb}$$
at $E_{\em lab}(\pi) = 300$ GeV
and the color-singlet prediction \cite{VHBT}
$$\sigma_{\rm singlet} (\pi N \into \psi' + X; x_F>0) = 3.2 \; {\rm nb}$$
gives an experimental upper limit of the color-octet contribution
(\ref{octetcontribution}). However, polarization measurements
show that all of the "missing" part cannot be due to
the lowest-order color-octet mechanisms.
The octet contribution is dominantly
transversely polarized; each of the three components alone
corresponds to $\alpha=1/2$, $\alpha=1$ and $\alpha=1$,
respectively, in
the angular distribution $1+\alpha\cos^2\theta$ of the $\psi'$
decay dileptons in the Gottfried-Jackson frame. On the
other hand, the observed
value is $\alpha=0.028 \pm 0.004$ \cite{Akerlof}.
Hence the color-octet mechanism could at most contribute about
half of the observed cross section, if the other half were
due to a mechanism which produces longitudinally polarized
$\psi'$ (note that the leading-twist color-singlet mechanism
gives $\alpha\approx 0.25$). We conclude that the leading-twist
color-singlet and color-octet mechanisms are not sufficient
to reproduce the observed polarization of the $\psi'$ in
pion-nucleon collisions.

In the color-singlet model, the ratio of the \jps and $\psi'$
cross sections is predicted to be
\beq
 \frac{\sigma(\psi')}{\sigma(J/\psi)}
 = \frac{M^3(J/\psi) \; \Gamma(\psi')}{M^3(\psi') \; \Gamma(J/\psi)}
 = 0.24,
 \label{CSMratio}
\eeq
where $M$ and $\Gamma$ are the masses and leptonic decay widths of
the \jps and $\psi'$. The experimental photoproduction
\cite{BinkleyPRL50} and fixed-target hadroproduction
\cite{Schuler} cross sections are consistent with
eq. (\ref{CSMratio}). This again suggests that the failure
of the existing models in reproducing fixed-target data
is due to neglected $\QQbar$ production
mechanisms rather than to the neglected color-octet
mechanisms of quarkonium formation.

Note that our results do not imply that the color-octet
explanation of the CDF quarkonium surplus is wrong. Using the
value \cite{ChoLeibovich}
$$
  <0|{\cal O}_8^{\psi'}(^3 S_1)|0> = 0.0073 \; {\rm GeV}^3,
$$
we find that the contribution from this non-perturbative transition
to $\sigma_{\rm octet}$ of eq. (\ref{octetcontribution}) is
only $0.7 {\rm \; nb} \; \times (1-\delta_{\lambda 0})$.
Such a small contribution, even though it is transversely
polarized, does not contradict the existing
fixed-target measurements.

\bigskip

The authors are grateful to S. J. Brodsky for important comments.
M.V. wishes to thank the theory group of SLAC for their warm
hospitality during his visit in January 1995, when the work reported
in this note was initiated.

\pagebreak

\bigskip \centerline{\large FIGURE CAPTIONS} \vspace{1cm}

{\bf Figure 1.}
Braaten and Fleming's color-octet fragmentation mechanism
of $\psi'$ production.

\bigskip

{\bf Figure 2.}
The Feynman diagrams which describe the lowest-order
color-octet $\psi'$ production mechanisms. We omit diagrams
that correspond to vanishing amplitudes.

\end{document}